\documentstyle[12pt]{article}
\begin{document}
\begin{center}

{\large \bf On Traversable Lorentzian Wormholes in the Vacuum
Low Energy Effective String Theory in Einstein and Jordan Frames}\\

\vspace{8mm}
K.K. Nandi \footnote{E-mail address: kamalnandi@hotmail.com}\\
  {\footnotesize \it Department of Mathematics, University of North Bengal,
    Darjeeling (W.B.) 734430, India\\
  Institute of Theoretical Physics, Chinese Academy of Sciences,
          P.O. Box 2735, Beijing 100080, China }\\

\vspace{4mm}

  Yuan-Zhong Zhang \footnote{E-mail address: yzhang@itp.ac.cn} \\

  {\footnotesize {\it CCAST(World Lab.), P. O. Box 8730, Beijing 100080, China \\
    Institute of Theoretical Physics, Chinese Academy of Sciences,
          P.O. Box 2735, Beijing 100080, China}}\\
\end{center}

\vspace{8mm}
\begin{abstract}
Three {\it new} classes (II-IV) of solutions of the vacuum low energy
effective string theory in four dimensions are derived. Wormhole solutions
are investigated in those solutions including the class I case both in the
Einstein and in the Jordan (string) frame. It turns out that, of the eight
classes of solutions investigated (four in the Einstein frame and four in
the corresponding string frame), massive Lorentzian traversable wormholes
exist in five classes. Nontrivial massless limit exists only in class I
Einstein frame solution while none at all exists in the string frame. An
investigation of test scalar charge motion in the class I solution in the
two frames is carried out by using the Pleba\'{n}ski-Sawicki theorem. A
curious consequence is that the motion around the extremal zero (Keplerian)
mass configuration leads, as a result of scalar-scalar interaction, to a new
hypothetical \textquotedblleft mass\textquotedblright\ that confines test
scalar charges in bound orbits, but does not interact with neutral test
particles.

\bigskip

\noindent PACS number(s): 04.40.Nr,04.20.Gz,04.62.+v
\end{abstract}

{\bf \vspace{10mm}}

\section{Introduction}

Currently, there exist an intense activity in the field of wormhole physics
following particularly the seminal works of Morris, Thorne and Yurtsever
[1]. Wormholes are topological handles that connect two distant otherwise
disconnected regions of space. Theoretical importance of such geometrical
objects is exemplified in several ways. For instance, they are invoked to
interpret or solve many outstanding issues both in the local as well as
cosmological scenarios [2-5]. Lorentzian wormholes could be threaded both by
quantum and classical matter fields that violate certain energy conditions
(``exotic matter") at least at the throat. In the quantum regime, several
negative energy density fields are already known to exist. For instance,
they occur in the Casimir effect, and in the context of Hawking evaporation
of black holes, and also in the squeezed vacuum states [1]. Classical fields
playing the role of exotic matter also exist. They are known to occur in the
$R+R^2$ theory [6], scalar tensor theories [7-11], Visser's cut and paste
thin shell geometries [12]. On general grounds, it has recently been shown
that the amount of exotic matter needed at the wormhole throat can be made
arbitrarily small thereby facilitating an easier construction of wormholes
[13].

A commendable arena to look for classical exotic fields is the vacuum linear
string theory which, in the low energy limit, reproduces a scalar tensor
theory of gravity in four dimensions. The action can be written in the
Jordan frame (JF), which is also called the string frame, and it is this
form of action that appears in the original nonlinear $\sigma $-model and
its solutions are what the string actually `sees'. (We set the $\beta $%
-functions to zero for reasons of quantum conformal invariance). The JF
action is referred to here as string theory. The action can also be cast
into the Brans-Dicke form with the coupling parameter $\omega =-1$ showing
that the Machian philosophy is already imbedded into the string action. This
Brans-Dicke action can be transferred to the conformally rescaled Einstein
frame\ (EF) so that the Lagrangian assumes the form of Einstein-Hilbert
action of (non-Machian) general relativity in which the scalar field
(dilaton) couples to the gravitional sector minimally but with an arbitrary
sign in the kinetic term. We choose to call the latter the Einstein massless
scalar (EMS) field theory. Both the signs can be theoretically allowed as
long as there does not appear any inconsistency. It should be noted that the
positive sign before the kinetic term in the action represents conventional
coupling while the negative sign corresponds to the unconventional one that
leads to the violation of energy conditions.

The motivation for the present paper is provided by three key reasons:
First, both the above frames exhibit certain symmetry properties, T-duality
in the Jordan frame and S-duality in the Einstein frame. Second, there is as
yet no consensus as to which frame is more physical although the Einstein
frame is often advocated in view of energy considerations. As we are here
concerned with only wormhole solutions, we need not be concerned with the
violations of energy conditions in either of the frames. Overall, there is
no canonical principle to rule out one frame in preference to the other and
hence we shall examine the solutions in both of them. This is the third
reason. In fact, recently, in the context of traversable Lorentzian
wormholes in general relativity Armend\'{a}riz-Pic\'{o}n (A-P) [14] has
shown that the most simple form of Lagrangian that satisfies all the
traversable wormhole conditions is that of EMS theory but with a {\it %
negative} sign before the kinetic term. The author has briefly
discussed massive wormholes in a certain class (let us call it
class I) of static, spherically symmetric solutions in the EMS
theory and has also proved the existence and stability of zero
mass wormholes. As remarked by the author, zero mass wormhole
configuration is the simplest and it exemplifies Wheeler's concept
of \textquotedblleft charge without charge". However, there also
exist other classes of EMS solutions although, unfortunately, they
have not received as much attention in the literature as the class
I solutions. Therefore, it is of interest to examine if wormholes
exist in those other classes of static spherically symmetric
solutions (II-IV) of EMS theory as well as in the string theory.
In this paper, we start with the EMS solutions from our earlier
paper and derive the corresponding new string frame solutions and
then adopt a search for wormholes analyzing all solutions on a
class by class basis in both the theories. The search result turns
out to be quite encouraging in the sense that out of the eight,
three and two classes represent massive wormhole solutions in the
EMS and string theory respectively. However, no massless limit
exists in the string theory. We shall also study the motion of
test particles in the gravity-scalar field
environment by adopting a different principle based on the Pleba\'{n}%
ski-Sawicki theorem. We work mainly in the Morris-Thorne
coordinate description for more transparency.

The paper is organized as follows: In Sec.2, we start from the linear string
action and review the class I solution of the EMS theory. In Sec.3, we
elucidate more pedagogical details of the EMS class I wormhole in coordinate
description and revisit the zero mass limit. The contents will be useful in
Sec.4 where we explore the wormhole nature of class I solution in the
context of string theory. Investigation of other classes of solutions
(II-IV) is contained in Sec.5. This section also includes the analyses of
the corresponding string classes of solutions. In Sec.6, we study test
particle motion in the class I solution of the two theories while in Sec.7,
we summarize our results. An appendix contains a comparison of notations for
easy reference.

\section{The action and class I solution: A brief review}

Our starting point is the 4-dimensional, low energy effective action of
heterotic string theory compactified on a 6-torus [15]. The tree level
string effective action, keeping only linear terms in the string tension $%
\alpha ^{\prime }$ and in the curvature $\widetilde{{\bf R}}$, takes the
following form in the ordinary-matter free region ($S_{matter}=0$):
$$
S_{eff}=\frac{1}{{\alpha ^{\prime }}}\int {d^{4}x\sqrt{-\tilde{g}}e^{-2%
\tilde{\Phi}}}\left[ \widetilde{{\bf R}}{+4\tilde{g}^{\mu \nu }\tilde{\Phi}%
_{,\mu }\tilde{\Phi}_{,\nu }}\right] ,\eqno(1)
$$%
where $\tilde{\Phi}$ is the dilaton field. Note that the zero values of
other matter fields do not lead to any additional constraints either on the
metric or on the dilaton [15]. One also avoids the complexity of abnormal
scalar coupling with these fields in the EMS version. Such couplings are
known to violate the principle of equivalence since the test particle rest
mass depends on the scalar field. We shall comment on this principle later
in Sec.6. Under the substitution $e^{-2\tilde{\Phi}}=\phi $, the above
action reduces to the JFBD action (we take the units $16\pi G=c=1$):
$$
S_{JF}=\int {d^{4}x\sqrt{-\tilde{g}}\left[ {\phi }\widetilde{{\bf R}}{+\frac{%
1}{\phi }\tilde{g}^{\mu \nu }\phi _{,\mu }\phi _{,\nu }}\right] }\eqno(2)
$$%
in which the BD coupling parameter $\omega $ is set to the value $\omega =-1$%
. This particular value is actually a model independent prediction
and it arises due to the fundamental symmetry of strings, viz.,
the target space duality [16]. It should be noted that the vacuum
BD action has a conformal invariance characterized by a constant
gauge parameter $\xi $. Arbitrary choice of $\xi $ can lead to a
shift from the value $\omega =-1$. This ambiguity can be removed
either by allowing abnormal coupling to matter or simply by fixing
the gauge [21]. We fix $\xi =0$. Under a further substitution
$$
g_{\mu \nu }^{\prime }=\phi \tilde{g}_{\mu \nu },\eqno(3)
$$%
$$
d\varphi ^{\prime }=\sqrt{\frac{2\omega +3}{2\alpha }} \frac{{d\phi }%
}{\phi },\quad \quad \alpha \neq 0,\eqno(4)
$$%
the action (2) goes into the EFBD action, for the string value
$\omega =-1$,
$$
S_{EF}=\int {d^{4}x\sqrt{-g^{\prime }}}\left[ {\bf R}{^{\prime
}+\alpha g^{\prime }{}^{\mu \nu }\varphi _{,\mu }^{\prime }
\varphi _{,\nu }^{\prime }}\right],   \eqno(5)
$$%
where we have introduced a constant arbitrary parameter $\alpha $
that can have any sign. The action (5) is also called the string
action in the Einstein frame but in this paper we distinguish it
as the action of the EMS theory. If the kinetic term $\alpha
g^{\prime }{}^{\mu \nu }\varphi _{,\mu }^{\prime }\varphi _{,\nu
}^{\prime }$ has an overall reverse (that is, negative) sign, we
have what one calls unconventional coupling. However, no matter
what the sign or value of $\alpha $ is, we can always proceed from
EMS action (5) backwards up to the string action (1). We keep
$\alpha $ unassigned until later. It seems remarkable that A-P
[14] has ended up with action (5) as the simplest action arguing
from a completely different angle, viz., by imposing wormhole
constraints on the Lagrangian for a general class of microscopic
scalar field. Obviously, the arguments have nothing to do with
string theory yet the end action is quite the same. So we have
here a picture in which the physics of dilatonic gravity meets
that of wormholes. In what follows, we shall use slightly
different notations that are in line with our earlier papers.
These can be easily transcribed to those in Ref.[14], as shown in
the appendix.

To clearly demarcate the scope of what follows, we must state that
we are not dealing here with time dependent cosmological
wormholes, and/or wormholes with Euclidean signatures [17-19]
which are qualitatively completely different from static
Lorentzian wormholes. However, the role of Eq.(4) that connects JF
and EF is the same. In this regard, note that we have imported a
new parameter $\alpha$ in Eq.(4) and it is obvious that the ranges
of $\omega$ and $\alpha$ can be chosen independently. In the
context of cosmology, the choice of $\omega =0$ leaves the
parameter $\alpha$ arbitrary in the EF [17,18]. It is also worth
noting that Quiros, Bonal and Cardenas [19] have shown that the
cosmological singularity occurring in the EF is removed in the JF
in the range $ - 3/2 < \omega  \le  - 4/3$. This result has
significant impact on the question of which frame, JF or EF, is
more physical and also on the status of quantum gravity [20].
However, in the context of string theory, we must use {\it only}
the model independent, unique string value $\omega =-1$ in Eq.(4).
In this case, we have $d\varphi ' = (1/\sqrt {2\alpha } )(d\phi
/\phi )$, and the range of $\alpha$ is essentially left
undetermined by the string theory field equations {\it per se} in
the EF, viz., Eqs.(6) and (7). What actually determines $\alpha$
is the condition for the existence of wormholes {\it at the
solution level} given, for instance, by $\beta ^2  > 1$ [see
Eq.(17) below], which in turn implies that $\varphi '$ be
imaginary for $\alpha >0$ [see Eqs.(11), (12)]. A-P [14] has shown
that the imaginary nature of $\varphi '$ does not lead to any
pathology or inconsistency in the physics of Lorentzian wormholes.
A completely equivalent but alternative description, again at the
solution level, is to regard $\varphi '$ as a real function which
then leads to $\alpha <0$. All these matters are developed in
Sections 2 and 3. The important point is that {\it both} the cases
[$\alpha >0$, $\varphi '$ imaginary, or, $\alpha <0$, $\varphi '$
real] lead to a negative sign before the kinetic term $\alpha
g'^{\mu \nu } \varphi '_{,\mu } \varphi '_{,\nu }$ which is what
we need for exotic matter. One is free to adopt any of the
mutually exclusive theoretical alternatives without any loss of
rigor in the wormhole analysis.

The field equations for the EMS theory, after dropping the primes in (5),
are given by
$$
R_{\mu \nu }=-\alpha \varphi _{, \mu } \varphi _{, \nu } \eqno(6)
$$
$$
\Box ^{2} \varphi =0. \eqno(7)
$$
In ``isotropic" coordinates $(x^\mu , \mu = 0,1,2,3)$, the
solution is given by [22]: \noindent
\[
ds^2 = g_{\mu \nu } dx^\mu dx^\nu \quad \quad \quad \quad \quad \quad \quad
\quad \quad \quad \quad \quad \quad \quad \quad \quad \quad \quad \quad
\quad \quad \quad \quad \quad \quad \quad \quad \quad \quad
\]
$$
\quad \quad \; = \left( {1 - \frac{m}{{2r}}} \right)^{2\beta } \left({1 +
\frac{m}{{2r}}} \right)^{ - 2\beta } dt^2 - \left( {1 - \frac{m}{{2r}}}
\right)^{2(1 - \beta )} \left( {1 + \frac{m}{{2r}}} \right)^{2(1 + \beta )} %
\left[ {dr^2 + r^2 d\Omega _2^2 } \right], \eqno(8)
$$

$$
\varphi (r)=2\lambda \ln \left[ {\frac{{1-\frac{m}{{2r}}}}{{1+\frac{m}{{2r}}}%
}}\right] ,\eqno(9)
$$%
$$
d\Omega _{2}^{2}=d\theta ^{2}+\sin ^{2}\theta d\phi ^{2},\eqno(10)
$$%
where $\beta ^{2}=1-2\alpha \lambda ^{2}$. This solution can be directly
obtained also by conformally rescaling the BD class I solution [9]. The two
undetermined constants $m$ and $\beta $ are related to the source strengths
of the gravitational and scalar parts of the configuration. To first order,
$$
\varphi \approx \frac{\sigma }{r},\eqno(11)
$$%
where
$$
\sigma =-2m\lambda =-2m\left[ {\left( {1-\beta ^{2}}\right) /2\alpha }\right]
^{1/2}\eqno(12)
$$%
is the strength of the scalar source. When $\beta =1$, we have $\varphi =0$,
and the Schwarzschild metric is recovered in accordance with the no hair
theorem. Using Einstein's energy momentum complex, we find that the total
mass $M$ of the configuration is given by
$$
M=m\beta .\eqno(13)
$$%
This is the conserved total mass of the configuration to be observed by
asymptotic observers. Using this value, the metric (8) can be expanded in
the weak field as
$$
ds^{2}=\left[ {1-2Mr^{-1}+2M^{2}r^{-2}+O(r^{-3})}\right] dt^{2}-\left[ {%
1+2Mr^{-1}+O(r^{-2})}\right] \left[ {dr^{2}+r^{2}d\Omega _{2}^{2}}\right] .%
\eqno(14)
$$%
This metric exactly coincides with the weak field Robertson
expansion [23] of a centrally symmetric gravitational field.
Assuming that the neutral test particles follow the geodesics
determined by the metric (8), that is no abnormal coupling of
ordinary matter with gravity, we see that all the well known solar
tests of gravity are described just as precisely as does the
exterior Schwarzschild metric. The parameter $\beta $ does not
appear separately in the expansion (14) and hence its effect can
not be measured by any metric test of gravity. The parameter
$\alpha $ does not appear here either but it does appear in the
expression for the scalar field in Eq.(9) or (11) and we can use
its sign to fix the nature of $\varphi $. Let us return to Eq.(13)
which can be immediately rewritten as
$$
M^{2}=m^{2}-\frac{1}{2}\alpha \sigma ^{2}.\eqno(15)
$$%
It is quite apparent from Eq.(12) that $\sigma $ can assume real or
imaginary values depending on the values assigned to $\beta $ and $\alpha $.
We shall now see what kind of values could be assigned to these parameters
if the metric (8) is to represent a traversable wormhole.

\section{Class I wormhole in the EMS theory}

For a coordinate description of wormholes that encapsules all the essential
details, the Morris-Thorne [1] form is most useful, which is given by
$$
ds^{2}=e^{2\Phi (R)}dt^{2}-\frac{1}{{1-\frac{{b(R)}}{R}}}dR^{2}-R^{2}d\Omega
_{2}^{2},\eqno(16)
$$%
where $\Phi (R)$ is the redshift function and $b(R)$ is the shape function.
Casting metric (8) in that form and manipulating a little, the wormhole
throat is found to occur at the isotropic $r$ coordinate radii
$$
r_{0}^{\pm }=\frac{m}{2}\left[ {\beta \pm \left( {\beta ^{2}-1}\right) ^{1/2}%
}\right] .\eqno(17)
$$%
The value $\beta ^{2}=1$ corresponds to a massive nontraversable wormhole
since $r_{0}^{\pm }$ coincides with the horizon radius $r_{s}=m/2$ and we
are not interested in this case. In order to build a traversable wormhole,
one needs to avoid this radius and therefore, one must have real $r_{0}^{\pm
}>m/2$. This requires that $\beta ^{2}>1$. Now consider scalar field energy
density $\rho $ and the Ricci scalar {\bf R} which work out to be
$$
\rho =\frac{1}{2}\times \frac{{m^{2}(1-\beta ^{2})}}{{\left( {1-\frac{{m^{2}}%
}{{4r^{2}}}}\right) ^{2}}}\times \left( {r+\frac{m}{2}}\right) ^{-2(1+\beta
)}\times \left( {r-\frac{m}{2}}\right) ^{-2(1-\beta )},\eqno(18)
$$%
$$
{\rm {\bf R}}=2m^{2}r^{4}\left( {1-\beta ^{2}}\right) \times \left( {r-\frac{%
m}{2}}\right) ^{-2(2-\beta )}\times \left( {r+\frac{m}{2}}\right)
^{-2(2+\beta )}.\eqno(19)
$$%
They become finite at $r=m/2$ if $\beta \geq 2$, which accords
well with the wormhole condition. In fact, it can be verified that
all curvature invariants are also finite under the condition
$\beta \geq 2$. So, one indeed has a regular spacetime, but the
problem is that the surface area becomes infinite at $r=m/2$. But
this could be due to a wrong choice of coordinates. Bronnikov {\it
et al }[24] called such spacetimes as representing cold black
holes (CBH) because of zero Hawking temperature. Some of their
interesting properties have also been discussed in the literature
[25]. In any case, the wormhole flares out to two asymptotically
flat regions connected by the throat and is traversable because
the tidal forces can be shown to be finite at the throat and
elsewhere.

For the wormhole value of $\beta ^{2}$, viz., $\beta ^{2}>1$, then, we have
two equivalent situations: (i) Take $\alpha <0$, say $\alpha =-2$. This
means breaking the energy conditions \textquotedblleft by hand" (we shall
provide an example later) in the source term in Eq.(6) so that we can have,
from Eq.(12), a real scalar charge $\sigma $, that is $\sigma ^{2}>0$, or
(ii) Take $\alpha >0$, say $\alpha =2$, then we have an imaginary scalar
charge from Eq.(12) so that $\sigma ^{2}=-\sigma ^{\prime }{}^{2}<0$. In
either case, of course, we have a reversed sign kinetic term in the action.
Also in Eq.(6), we have a stress tensor that violates all energy conditions
giving the kind of classical exotic matter necessary for the threading of
traversable wormholes. Then, from Eq.(15), we have
$$
M^{2}=m^{2}+\sigma ^{\prime }{}^{2}.\eqno(20)
$$%
A wormhole with zero total mass, that is, $M=0$, immediately implies $m=0$
and $\sigma ^{\prime }=0$. In other words, we have the trivial case of a
flat metric and zero scalar field. However, it is possible to avoid this
uninteresting case by making $m$ also imaginary and noting from Eq.(13) that
we can also have $M=0$ if we set $\beta =0$. This is actually the case
considered by A-P [14]. In fact, taking $\alpha =2,\beta =0$ we have from
Eq.(15),
$$
\sigma ^{2}=m^{2}.\eqno(21)
$$%
Clearly, if $\sigma ^{2}$ is negative, then so is $m^{2}$ and vice versa. It
is thus enough in this particular case to assume the imaginary nature of any
{\it one} of them. Defining the proper distance $l$ as
$$
l=r-\frac{{\sigma ^{\prime }{}^{2}}}{{4r}},\eqno(22)
$$%
the metric (8) can be rewritten in the form
$$
ds^{2}=dt^{2}-dl^{2}-\left( {l^{2}+\sigma ^{\prime }{}^{2}}\right) d\Omega
_{2}^{2},\eqno(23)
$$%
$$
\varphi =\ln \left[ {\frac{{1-\frac{{i\sigma ^{\prime }}}{{2r(l)}}}}{{1+%
\frac{{i\sigma ^{\prime }}}{{2r(l)}}}}}\right] ,\eqno(24)
$$%
$$
r(l)=\frac{{l\pm \sqrt{l^{2}+\sigma ^{\prime }{}^{2}}}}{2}.\eqno(25)
$$%
Since $m^{2}$ is also negative, i.e., $m^{2}=-m^{\prime }{}^{2}<0$, the
wormhole throat at $l=0$ implies the real coordinate values $r_{0}^{\pm
}=\pm \frac{{\sigma ^{\prime }}}{2}=\mp \frac{{m^{\prime }}}{2}$ and the
scalar field $\varphi $ becomes imaginary but does not blow up at this
value. Also, in the units considered, we have at $r=r_{0}^{\pm }$, $\rho =-%
\frac{1}{{2\sigma ^{\prime }{}^{2}}}$, {\bf R} $=-\frac{2}{{\sigma ^{\prime
}{}^{2}}}$. Thus, we indeed have the simplest well behaved wormhole. Under
the considerations above, we now have a real equation that Eq.(20)
translates into, viz., $M^{2}=-m^{\prime }{}^{2}+\sigma ^{\prime }{}^{2}$
obviously implying that $M=0$ wormholes are {\it extremal} in nature. This
amounts to saying that we have a configuration in which the stresses of the $%
\varphi $ field contribute an amount of energy just sufficient to nullify
the effect of gravitational potential making the total energy zero. In other
words, we have nontrivial energy sources residing at the origin of central
symmetry in such a way as to make a configuration that is gravitationally
indifferent to neutral test particles. Note that the extremal configuration
can arise even when no exotic matter is involved, that is, $\beta ^{2}<1$.
In this case also, we can have $M=0\Rightarrow m=\sigma $ from Eq.(15)
simply by choosing $\alpha =2$. The foregoing analyses will be helpful in
what follows.

Finally, it must be noted that class I EMS solutions have received
good attention in the literature [26, 27]. For instance, using
Eqs. (23)--(25), particle models in general relativity have been
constructed by Ellis [26] by way of an ether flow through a
drainhole. Geometrical optics, classical and quantum scattering
problems have been studied in the Ellis geometry by Chetouani and
Clement [28] and by Clement [29].

\section{Class I wormhole in the string theory}

Starting with the solutions (8) and (9) and working backwards up to action
(1), we can straightaway write down the corresponding string solution as
\[
d\tilde{s}^{2}=\tilde{g}_{\mu \nu }dx^{\mu }dx^{\nu }\quad \quad \quad \quad
\quad \quad \quad \quad \quad \quad \quad \quad \quad \quad \quad \quad
\quad \quad \quad \quad \quad \quad
\]%
\[
=\left( {1-\frac{m}{{2r}}}\right) ^{2(\beta -\lambda \sqrt{2\alpha })}\left(
{1+\frac{m}{{2r}}}\right) ^{-2(\beta -\lambda \sqrt{2\alpha })}dt^{2}\quad
\quad \quad \quad \quad \quad \quad
\]%
$$
-\left( {1-\frac{m}{{2r}}}\right) ^{2(1-\beta -\lambda \sqrt{2\alpha }%
)}\left( {1+\frac{m}{{2r}}}\right) ^{2(1+\beta +\lambda \sqrt{2\alpha })}%
\left[ {dr^{2}+r^{2}d\Omega _{2}^{2}}\right] ,\eqno(26)
$$%
$$
\tilde{\Phi}=-\lambda \sqrt{2\alpha }\ln \left[ {\frac{{1-\frac{m}{{2r}}}}{{%
1+\frac{m}{{2r}}}}}\right] .\eqno(27)
$$%
Under a suitable re-identification of constants and coordinates, the above
solution reduces to that discussed by Kar [15]. To first order in $(1/r)$,
we have the strength of the dilatonic source $\tilde{\sigma}$ given by
$$
\tilde{\Phi}\approx \frac{{\tilde{\sigma}}}{r},\quad \tilde{\sigma}=m\lambda
\sqrt{2\alpha }.\eqno(28)
$$%
One recovers the Schwarzschild metric in the limit $\beta =1$. However, it
does not seem possible to recover the seed solutions (8) and (9) which shows
that we are dealing here with a class of solutions that is essentially
distinct from its counterpart either in BD or in the EMS theory.

Expanding the metric (26) as in (14), and identifying the Keplerian mass $%
M^* $, the tensor mass $M_T =m\beta$ and the scalar mass $M_S = -
m\sqrt {1 - \beta ^2 }$ (cf. [30] for these definitions), we have
\[
d\tilde s^2 = \left[ {1 - 2M^* r^{ - 1} + 2M^{*2} r^{ - 2} + O(r^{ - 3} )} %
\right]dt^2 \quad \quad \quad \quad \quad \quad \quad \quad \quad \quad
\quad
\]
$$
- \left[ {1 + 2M^* r^{ - 1} \left( {\frac{{\beta + \sqrt {1 - \beta ^2 } }}{{%
\beta - \sqrt {1 - \beta ^2 } }}} \right) + O(r^{ - 2} )} \right]\left[ {%
dr^2 + r^2 d\Omega _2^2 } \right],     \eqno(29)
$$
where $M^* = m\left( {\beta - \sqrt {1 - \beta ^2 } } \right)
\equiv M_T + M_S$. Solar observations can put a limit to $\beta$,
which obviously is expected to be $\beta \approx 1$. Note that the
motion of ordinary test particle measures the Keplerian mass and
it is assumed that the particle has negligible self energy so that
the Nordvedt effect can be ignored. The tensor mass is measured by
the motion of a Schwarzschild black hole in the metric (8)[30].

We shall now investigate if the solutions (26) and (27) represent
traversable wormholes. To this end, we cast the metric (26) in the
Morris-Thorne form by redefining the radial variable $r\rightarrow R$ as
$$
R=r\left( {1-\frac{m}{{2r}}}\right) ^{(1-\beta -\lambda \sqrt{2\alpha }%
)}\left( {1+\frac{m}{{2r}}}\right) ^{(1+\beta +\lambda \sqrt{2\alpha })}.%
\eqno(30)
$$%
The redshift function $\Phi (R)$ and the shape function $b(R)$ turn out to
be
$$
\Phi (R)=\left[ {\beta -\sqrt{1-\beta ^{2}}}\right] \times \left[ {\ln
\left( {1-\frac{m}{{2r}}}\right) -\ln \left( {1+\frac{{m}}{{2r}}}\right) }%
\right] ,\eqno(31)
$$%
$$
b(R)=R\left[ {1-\left( {\frac{{r^{2}+\frac{{m^{2}}}{4}-m\tilde{\beta}r}}{{%
r^{2}-\frac{{m^{2}}}{4}}}}\right) ^{2}}\right] ,\eqno(32)
$$%
$$
\tilde{\beta}\equiv \beta +\sqrt{1-\beta ^{2}}.\eqno(33)
$$%
Clearly, $\beta =1\Rightarrow \tilde{\beta}=1$. The throat occurs at the
radii
$$
\tilde{r}_{0}^{\pm }=\frac{m}{2}\left[ {\tilde{\beta}\pm \left( {\tilde{\beta%
}^{2}-1}\right) ^{1/2}}\right] \eqno(34)
$$%
and as before the wormhole requirement is that $\tilde{\beta}^{2}>1$. The
energy density $\tilde{\rho}$ is given by
$$
\tilde{\rho}=\frac{2}{{R^{2}}}\times \frac{{m^{2}r^{2}}}{{(r^{2}-m^{2}/4)^{2}%
}}\times (1-\tilde{\beta}^{2}),\eqno(35)
$$%
which is negative for $\tilde{\beta}^{2}>1$, as expected. The tidal forces
are finite and so the solution represents a traversable wormhole. The total
conserved mass of the configuration as observed by asymptotic observers can
be identified as
$$
\tilde{M}=m\tilde{\beta}.\eqno(36)
$$

In the zero mass limit: $\tilde{\beta}=0\Rightarrow \tilde{M}=0$. This
implies $\beta =-1/\sqrt{2}$ which in turn implies a dilatonic field
strength $\tilde{\sigma}=m/\sqrt{2}$. It is now useful to recall the
discussions surrounding Eqs.(21)-(25). The situation here is that ${\tilde{%
r_{0}}}^{\pm }$ is imaginary as before but to make it real one can assume $m$
to be imaginary which automatically implies that $\tilde{\sigma}$ is also
imaginary. Under the transformation $l=r\left( {1+\frac{{m^{2}}}{{4r^{2}}}}%
\right) $, the metric (26) becomes
$$
d\tilde{s}^{2}=\left( {\frac{{l+m}}{{l-m}}}\right) ^{\sqrt{2}%
}dt^{2}-dl^{2}-\left( {l^{2}-m^{2}}\right) d\Omega _{2}^{2},\quad \quad
\tilde{\Phi}=-\frac{1}{{2\sqrt{2}}}\ln \left( {\frac{{l-m}}{{l+m}}}\right) .%
\eqno(37)
$$%
With $m$ imaginary, we have a positive definite minimum surface area $-4\pi
m^{2}$ but at $l=0$, we have $\tilde{g}_{00}=(-1)^{\sqrt{2}}$ which is a
many valued function. Also for $l\neq 0$, $\tilde{g}_{00}$ becomes imaginary
which requires us to go beyond the real manifold that we have been
considering. Hence, although massive wormholes exist, zero mass wormholes
seem untenable, at least in the simplest form of the string theory that is
under present investigation. We shall also state a physical reason in Sec.6
as to why they are untenable. Nonetheless, (37) is still a {\it formal} zero
mass solution of the string action (1). We shall encounter solutions of
similar nature in the next section.

The developments in Sections 3 and 4 immediately reveal certain interesting
features about the images of EMS class I wormholes in the string theory. It
follows that both $\beta =0$ (zero mass) and $\beta =1$ EMS wormholes have
the image of only a nontraversable wormhole in the string theory with the
throat occurring at $\tilde{r}_{0}^{\pm }=m/2$. For $\beta =0$ we have $%
\tilde{g}_{00}\rightarrow \infty $ at $\tilde{r}_{0}^{\pm }=m/2$, which is
undesirable. For the range of values $\beta ^{2}>1$, traversable Lorentzian
wormholes do exist in the EMS theory but they have no counterpart in the
string theory since $\tilde{\beta}$ and $\tilde{r}_{0}^{\pm }$ become
imaginary. However, ordinary EMS solutions for $\beta ^{2}<1$ have {\it only}
wormhole images in the string regime due to the fact that $\beta
^{2}<1\Rightarrow \tilde{\beta}^{2}>1$ which was shown to be a necessary
condition for string wormholes.

\section{Class II-IV solutions in the EMS and string theory}

By conformal rescaling of the BD class II-IV solutions, it is
possible to obtain the corresponding solutions in the EMS theory
[31]. Alternatively, they can be obtained by solving the EMS
equations (6) and (7) by standard procedures. We take the general
form of the metric as
$$
ds^{2}=P(r)dt^{2}-Q(r)\left[ {dr^{2}+r^{2}d\Omega _{2}^{2}}\right] .\eqno%
(38)
$$
\ \ \ \ \ (a) Class II EMS solutions are
$$
P(r)=e^{2\alpha _{0}+4\gamma \arctan \left( {r/b}\right) }\quad \quad Q(r)=%
\left[ {1+\frac{{b^{2}}}{{r^{2}}}}\right] ^{2}e^{2\beta _{0}-4\gamma \arctan
\left( {r/b}\right) },\eqno(39)
$$%
$$
\varphi (r)=2\lambda \arctan \left( {\frac{r}{b}}\right) ,\eqno(40)
$$%
where $\lambda \equiv \left[ {\frac{{\left( {1+\gamma ^{2}}\right) }}{{%
2\alpha }}}\right] ^{1/2}$ and $\alpha _{0},\beta _{0},\gamma ,b$
are arbitrary constants. Asymptotic flatness requires that $\alpha
_{0}=-\pi \gamma ,\;\beta _{0}=\pi \gamma $. The solution (39) has
a conserved total energy $M=2b\gamma $ as can be verified by
computing the Einstein complex of stress energy. With this value
of $M$, the metric expands exactly like Eq.(14) for $r^{2}\geq
b^{2}$, and thereby explains all the solar system tests of
gravity. To see if the solution set (38)-(40) represent wormholes,
we cast it in the Morris-Thorne form to find that the coordinate
throat radii occur at
$$
r_{0}^{\pm }=b\left[ {\gamma \pm \sqrt{1+\gamma ^{2}}}\right] .\eqno(41)
$$%
Note that, in the solution (40), one has $1+\gamma ^{2}<0$, and this
inequality is a result of the field equations (8) and (9), so that one has
an imaginary $\varphi $. Alternatively, we can choose $\alpha <0$, and have
a real $\varphi $. But this is no real problem as the two situations are
equivalent, as explained earlier. Although $r_{0}^{\pm }$ is imaginary, one
might choose $b$ to be imaginary to make $r_{0}^{\pm }$ real. In this case, $%
M$ also becomes real. As is obvious, we are employing the same
arguments as we did in Sec.3 for the zero mass case. Note that,
although $\gamma $ and $b$ are imaginary, the metric functions
$P(r)$ and $Q(r)$ are real. All the curvature invariants are
finite everywhere [31] and the tidal forces experienced by a
geodesic traveler can be shown to be finite at the wormhole throat
and these tend to vanish at the asymptotic region. Most
importantly, the exponential function $P(r)$ does not vanish
anywhere so that the solution has no horizon. In this sense, it
shares the features of Morris-Thorne \textquotedblleft $\Phi
=0$\textquotedblright\ (no horizon) wormholes [1]. Thus, the EMS
class II solutions also represent traversable wormholes. Although
in the zero mass limit, viz., $\gamma =0$, the metric (38) in
proper distance language looks promising, that is,
$$
ds^{2}=dt^{2}-dl^{2}-\left( {l^{2}+4b^{2}}\right) d\Omega _{2}^{2},\eqno(42)
$$%
such wormholes unfortunately do not exist as the limit itself ($\gamma =0$)
conflicts with the inequality $1+\gamma ^{2}<0$.

The string version of the class II solution is given by
$$
d\tilde{s}^{2}=\tilde{P}(r)dt^{2}-\tilde{Q}(r)[dr^{2}+r^{2}d\Omega _{2}^{2}],%
\eqno(43)
$$%
$$
\tilde{P}(r)=e^{2\alpha _{0}+4[\gamma -\sqrt{1+\gamma ^{2}}]\arctan \left( {%
r/b}\right) },\quad \tilde{Q}(r)=\left[ {1+\frac{{b^{2}}}{{r^{2}}}}\right]
^{2}e^{2\beta _{0}-4[\gamma +\sqrt{1+\gamma ^{2}}]\arctan \left( {r/b}%
\right) },\eqno(44)
$$%
$$
\tilde{\Phi}(r)=-(\gamma -\sqrt{1+\gamma ^{2}})\arctan \left( {\frac{r}{b}}%
\right) .\eqno(45)
$$%
Other relevant quantities, e.g., the mass and throat radii, are respectively
given by
$$
\tilde{M}=2b\tilde{\gamma}=2b\left[ {\gamma +\sqrt{1+\gamma ^{2}}}\right]
,\quad \tilde{r}_{0}^{\pm }=b\left[ {\tilde{\gamma}\pm \left( {1+\tilde{%
\gamma}^{2}}\right) ^{1/2}}\right] .\eqno(46)
$$%
All these quantities are real if $b$ is imaginary, as we assumed. Here
again, for the same reasons described in Sec.4, massive, i.e., $\tilde{M}%
\neq 0$ traversable wormholes do exist but the massless limit does not, as
no value of $\gamma $ can make $\gamma +\sqrt{1+\gamma ^{2}}=0$, a condition
that is required to make $\widetilde{M}=0$.

(b) Class III EMS solutions are
$$
P(r)=\alpha _{0}e^{-\left( {\gamma r/b}\right) },\quad \;Q(r)=\beta
_{0}\left( {\frac{r}{b}}\right) ^{-4}e^{\left( {\gamma r/b}\right) },\quad
\;\varphi (r)=\frac{{\gamma r}}{{2b}}.\eqno(47)
$$

This solution is not asymptotically flat and hence does not meet the
requirement of asymptotic flaring out of the wormholes. However, it is flat
in the limit $r \to 0$. If we still formally impose the zero mass condition $%
\gamma =0$ and define $l=-b^2 /r$, we have the metric
$$
ds^2 = \alpha _0 dt^2 - \beta _0 dl^2 -\beta _0 l^2 d\Omega _2^2 . \eqno(48)
$$
Under a further rescaling $\sqrt {\alpha _0 } t \to t^{\prime}$, $\sqrt {%
\beta _0 } l \to l^{\prime}$, we end up with a trivial metric. The string
class III metric is
$$
\tilde P(r) = \alpha _0 e^{ - \left( {3\gamma r/2b} \right)} ,\;\quad \tilde
Q(r) = \beta _0 \left( {\frac{r}{b}} \right)^{ - 4} e^{\left( {\gamma r/2b}
\right)} ,\;\quad \tilde \Phi (r) = - \frac{{\gamma r}}{{2b}}, \eqno(49)
$$
and we again have a flat Minkowski metric like Eq.(48) for the case $\gamma
=0$.

(c) Class IV EMS solutions exhibit some good properties. They are given by
$$
P(r)=\alpha _{0}e^{-\left[ {\gamma /(br)}\right] },\;\quad Q(r)=\beta _{0}e^{%
\left[ {\gamma /(br)}\right] },\;\quad \varphi (r)=-\frac{\gamma }{{2br}},%
\eqno(50)
$$%
Asymptotic flatness fixes $\alpha _{0}=\beta _{0}=1$. The horizon
appears at $r=0$. First of all, under the transformation
$r\rightarrow 1/r$, $\beta _{0}\rightarrow b^{4}\beta _{0}$, the
solution goes over to the class III EMS solution (47) and hence
the two classes are not essentially distinct, although in the
original JF version, they are cited as different classes of
solutions [32]. Secondly, all the curvature invariants are finite
everywhere including $r=0$. In fact, the solution could be
interpreted as a CBH [25] since the area at $r=0$ is infinite. But
the geodesic congruences can not reach the origin, but reach a
minimum distance $r_{0}=M$ (see below) away from it, corresponding
to a finite surface area. Thereafter, they diverge so that it is
more likely that it represents a pure wormhole. Thirdly, the total
conserved mass for the solution is given by a real $M=\gamma /2b$
and the metric exactly coincides with the expansion (14) up to the
orders considered. Hence, it describes all the weak field tests of
general
relativity just as good as the Schwarzschild metric does. However, for $%
\gamma =0$ for which $M=0$, we again have only a flat spacetime and a
vanishing scalar field and consequently no zero mass wormholes. In the
scalar field theory, there is a black hole counterpart which usually occurs
when the scalar field is set to zero in accordance with the "no hair"
theorem. This situation obviously does not arise here and that is another
reason why we prefer to call class IV solutions as pure wormholes.

To see if class IV EMS solutions represent massive wormholes, we cast the
metric (50) in the Morris-Thorne form to obtain the shape and redshift
functions, respectively, as
$$
b(R)=re^{\frac{M}{r}}\left[ {1-\left( {1-\frac{M}{r}}\right) ^{2}}\right]
,\quad \quad \Phi (R)=-\frac{M}{r},R=re^{\frac{M}{r}}.\eqno(51)
$$%
The wormhole throat appears at $r_{0}=M\Rightarrow R=Me$ which is greater
than the horizon radius. The density $\rho $ and the radial tension $\tau $
are
$$
\rho =-\frac{{M^{2}}}{{R^{2}r^{2}}}<0,\quad \quad \tau =\frac{{M^{2}}}{{%
R^{2}r^{2}}},\eqno(52)
$$%
such that $\tau -\rho >0$ not only at the throat but everywhere. Hence the
flaring out condition is satisfied. The tidal forces are finite for static
as well as for freely falling observers [21]. The forces, however, could be
large for small values of $M$. So, everything put together, the solution
indeed represents a massive Lorentzian wormhole that is traversable at least
in principle.

Note that the solution (50) solves the field equations (8) and (9) for $%
\alpha =-2$ so that here we have an example where all energy conditions are
broken by hand, since $\varphi $ is real. In order to go to the string
metric, we need $\phi $ which becomes, with this value of $\alpha $, $\phi
^{-1}=e^{-2i\varphi }$. This imports an imaginary factor to the string
metric. Therefore, the string version of class IV solution has the form
$$
\tilde{P}(r)=e^{-\frac{{2M}}{r}\left( {1-i}\right) },\quad \quad \tilde{Q}%
(r)=e^{\frac{{2M}}{r}\left( {1+i}\right) },\quad \quad \tilde{\Phi}=\frac{{iM%
}}{r}.\eqno(53)
$$%
Clearly, Eqs.(53) do not represent wormholes in real spacetimes although it
is an interesting formal solution of string field equations of the action
(1) in the same way as the zero mass solution, Eq.(37) is. Here again, $%
\tilde{M}=M(1+i)=0\Rightarrow M=0$, and this leads to a trivial Minkowski
spacetime so that there are no zero mass wormholes.

\section{Charged test particle motion}

In this section, we consider motion only in the class I solution of EMS and
string theory. As mentioned in Sec.2, from the expansion (14) the effect of $%
\beta $ (or the source scalar charge $\sigma $) can not be separately
explored by the motion of neutral test particles at least in the power of $%
1/r^{2}$ since the total conserved gravitating mass $M$ appears only as a
product of $m$ and $\beta $. The expansion (29) in the string version does
separate the effect of $\beta $ and neutral test particle probes are able to
put a limit on its value. However, here we wish to consider not neutral but
charged particle motions. To this end, following an interesting approach by
Buchdahl [22], we regard $\varphi $ (or in the string context, $\tilde{\Phi}$%
) as representing some medium or long range force field existing
in spacetime $g_{\mu \nu }$ (or $\tilde{g}_{\mu \nu }$) and
imagine a test scalar charge responding to this field directly in
addition to interacting indirectly via the metric. The situation
is analogous to the motion of an electrically charged particle in
the Reissner-Nordstr\"{o}m spacetime. In order to have bound
orbits, we shall assume that the source and test charges have
opposite signs. With this understanding, let us consider the
equation of motion of a test particle with infinitesimally small
mass $\delta $ and scalar charge $\varepsilon $. In virtue of the
Pleba\'{n}ski-Sawicki theorem [33], the geodesic equations are
given by
$$
u^{\mu }\left[ {\left( {\delta -\varepsilon \varphi }\right) u^{\nu }}\right]
_{;\mu }=-\varepsilon \varphi _{;}^{\nu },\eqno(54)
$$%
where $u^{\mu }$ is the four velocity of the particle and
$;$denotes covariant derivative with respect to $g_{\mu \nu }$
defined in (8). These equations have the first integral $u_{\mu
}u^{\mu }=1$ and the particle trajectories correspond to those
defined by the metric [22]
$$
ds^{\prime }{}^{2}=\left( {\delta -\varepsilon \varphi }\right) ^{2}g_{\mu
\nu }dx^{\mu }dx^{\nu }.\eqno(55)
$$%
By carrying out the expansion plugging in the expressions of $g_{\mu \nu }$
and $\varphi $ from (8) and (9), we have the metric
$$
ds^{\prime }{}^{2}=h^{\prime }(r)dt^{2}-p^{\prime }(r)\left[ {%
dr^{2}+r^{2}d\Omega _{2}^{2}}\right] ,\eqno(56)
$$%
where
$$
h^{\prime }(r)=1-2(1-\varpi )\frac{M}{r}+\left( {2-4\varpi +\varpi ^{2}}%
\right) \frac{{M^{2}}}{{r^{2}}}+O(\frac{1}{{r^{3}}}),\eqno(57)
$$%
$$
p^{\prime }(r)=1+2\left( {1+\varpi }\right) \frac{M}{r}+O(\frac{1}{{r^{2}}}%
),\quad \quad \varpi =\frac{{\sigma \varepsilon }}{{\delta M}},\eqno(58)
$$%
in which $\varpi $ can take on only negative values as $\epsilon $ and $%
\sigma $ are assumed to have opposite signs. All the observable quantities
relating to the trajectory of the test scalar charge can be calculated in
the usual way. For instance, if $\kappa $ is the precession of the
pericenter per revolution for a given $\varpi $ and $\kappa _{0}$ is the
precession for $\varpi =0$, then
$$
\frac{\kappa }{{\kappa _{0}}}=1-\frac{{2\varpi }}{3}-\frac{{\varpi ^{2}}}{6}.%
\eqno(59)
$$%
Note that $\kappa =\kappa _{0}\Rightarrow \varpi =-4$. Now, in the
environment of a $M=0$ extremal configuration, we immediately find that the
metric functions reduce to
$$
h_{0}^{\prime }(r)=1-\frac{{2M^{\prime }}}{r}+\frac{{M^{\prime }{}^{2}}}{{%
r^{2}}}+O(\frac{1}{{r^{3}}}),\eqno(60)
$$%
$$
p_{0}^{\prime }(r)=1+\frac{{2M^{\prime }}}{r}+O(\frac{1}{{r^{2}}}),\eqno(61)
$$%
that can be thought of as generated by a hypothetical scalar
\textquotedblleft mass"
$$
M^{\prime }=\frac{{\sigma \varepsilon }}{\delta }>0.\eqno(62)
$$%
Although neutral test particles follow straight paths due to the fact that
the Keplerian source mass $M=0$, the test scalar charge executes a motion
that closely resembles that of a neutral test particle in the Schwarzschild
spacetime generated by the mass $M^{\prime }=\sigma \varepsilon /\delta $.
There will of course be a slight difference in the numerical value of the
precession of orbits due to the lack of factor 2 in the $(1/r^{2})$ term in
Eq.(60). Nevertheless, what we have here is a purely scalar-scalar
interaction leading to a scalar mass $M^{\prime }$ that restrains the test
charges in their geodesics but does not respond gravitationally. In the
limit $\delta \rightarrow 0,M^{\prime }$ could be very large. This is an
interesting feature of string theory if we believe that physics is described
by the EMS action (5). One finds that the scalar-scalar interaction in an
otherwise flat space (as noticed by neutral particles) describes a kind of
confinement of the test scalar charge in bound orbits.

In the string environment, $\varphi $ should be replaced by $\tilde{\Phi}$
and $g_{\mu \nu }\rightarrow \tilde{g}_{\mu \nu }$, and then the counterpart
of the metric (56) becomes
$$
d\tilde{s}^{\prime }{}^{2}=\left( {\delta -\varepsilon \tilde{\Phi}}\right)
^{2}d\tilde{s}^{2}=\tilde{h}(r)dt^{2}-\tilde{p}(r)\left[ {%
dr^{2}+r^{2}d\Omega _{2}^{2}}\right] .\eqno(63)
$$%
Using the metric functions given by (29), we have
$$
\tilde{h}(r)=1-2(1-\varpi _{1})\left( {\frac{{M^{\ast }}}{r}}\right) +\left(
{2-4\varpi _{1}+\varpi _{1}^{2}}\right) \left( {\frac{{M^{\ast }}}{r}}%
\right) ^{2}+O\left( {\frac{1}{{r^{3}}}}\right) ,\eqno(64)
$$%
$$
\tilde{p}(r)=1+2\left[ {\left( {1+\varpi _{1}}\right) \left( {\frac{{\beta +%
\sqrt{1-\beta ^{2}}}}{{\beta -\sqrt{1-\beta ^{2}}}}}\right) }\right] \frac{{%
M^{\ast }}}{r}+O\left( {\frac{1}{{r^{2}}}}\right) ,\quad \quad \varpi _{1}=%
\frac{{\varepsilon \tilde{\sigma}}}{{\delta M^{\ast }}}.\eqno(65)
$$%
For $\tilde{M}=0$ ordinary string configuration (not wormhole, they do not
exist as shown in Sec.4) corresponding to $\tilde{\beta}=0$, we find that
the term in the square bracket vanishes identically. Therefore, the motion
in this situation would be somewhat different from the one described by
Eqs.(56)-(58). If in addition, we assume that the Keplerian mass $M^{\ast }=0
$ in Eq. (64), we have the metric function $\tilde{h}(r)$ analogous to
Eq.(62) with $M^{\prime }$ replaced by $\frac{{\varepsilon \tilde{\sigma}}}{%
\delta }$, but a flat $\tilde{p}(r)$ up to first order. Thus, in a flat
metric with only the dilatonic field coupling to it, there is still some
sort of a dilaton-dilaton interaction generating a certain mass if the
physics is believed to be described by the action (1).

The developments of this section have a direct bearing on the
principle of equivalence that is usually discussed in terms of
motion of the test particles in a given metric. If the motion
follows the geodesics of the metric, one says that the principle
is obeyed. First consider the BD theory. The weak principle of
equivalence (WEP) is satisfied as small (negligible binding
energy) neutral test particles do move along the geodesics
determined by the BD metric. But the strong equivalence principle
(SEP), which states that WEP must be satisfied even for bodies
with large gravitational self-energy, is violated in the BD theory
due to the appearance of two types of masses [30]. In the string
theory, too, WEP is satisfied to the same extent as in BD theory,
but we see that SEP could be violated because of the appearance of
two masses $M_T $ and $M_S $ in the metric (26). The ratio of the
two masses, viz., $M_S /M_T = (\sqrt {1- \beta ^2 })/\beta $
depends on the gravitational binding energy of the source. Indeed,
in the limit $\tilde\beta\to 0$, one has $\left| {M_S /M_T }
\right| \to 1$, in contrast to the EMS case in which one has $M_S
=0$ . Thus, in the string zero mass configuration, the self energy
becomes maximum. This is probably an indicator as to why zero mass
wormholes in the string theory do not exist.

In this context, recall that the metric $g_{\mu \nu }$ (it is also
called the Pauli metric) couples to dilaton in a \textquotedblleft
normal" way, that is, by way of EMS action (5) and it has been
argued that the dilatonic test particle (and not the ordinary
neutral paricle) should follow the geodesics of $g_{\mu \nu }$ and
satisfy the WEP [21]. If we endow the dilaton with an
infinitesimal mass $\delta $ (e.g., pseudo Goldstone bosons) and
charge $\varepsilon $, then the argument seems to be at variance
with the Pleba\'{n}ski-Sawicki theorem due to the fact that the
dilaton follows the geodesics of $ds^{\prime }{}^{2}$ and {\it
not} of $ds^{2}$. Indeed, a value of $\kappa /\kappa _{0}$ away
from unity indicates a violation of WEP by charged particle
motion. However, this need not necessarily be the case. One may
always take the usual viewpoint that the effect of the source
scalar field is already subsumed in the metric and the motion of
test dilaton, by definition, does not alter the background
geometry. It is only if one allows a departure from this viewpoint
by introducing an extra scalar-scalar interaction {\it a la}
Buchdahl that one comes up with what is embodied in the Eqs.
(60)-(65).

\section{Summary}

The contents of the paper may be summarized as under:

(1) Using the four classes (I-IV) of EMS solutions [action (5)] as
seed solutions derived earlier [31], corresponding classes of
solutions in the low energy effective string theory [action (1)]
have been obtained and analyzed. They can be grouped as string
class I [Eqs.(26), (27)], class II [Eqs. (44), (45)], class III
[Eq.(49) and class IV [Eq. (53)] static spherically symmetric
solutions. The string class I solutions can be identified with
those discussed recently by Kar [15], but the rest of the three
sets of solutions given here are essentially {\it new} to our
knowledge. Note that it is not possible to recover the seed EMS
solutions by any choice of the free parameters. The metric parts
of the string solutions resemble those in the BD theory with
$\omega =-1$, while the scalar parts correspond to those of EMS
theory up to a redefinition of constants. In other words, the
solutions have one leg in the BD and other in the EMS theory.
These solutions could be derived also by solving the whole
plethora of string field equations coming from the action (1). The
last solution set, Eqs.(53), is complex but that is not unexpected
as the seed BD class IV solution is also complex at the string
value $\omega =-1$. It should be emphasized that the solutions
discussed here do not exhaust all the possible spherically
symmetric solutions of the string theory that might exist.

(2) Wormhole solutions have been explored in all the above solutions of EMS
and string theories. The following results are obtained: Massive Lorentzian
traversable wormholes exist in classes I, II and IV in the EMS theory, but
not in class III. However, it has been pointed out that class III solution
is related to class IV through a coordinate inversion and are not really
distinct solutions. However, for the wormhole flaring out condition, the
latter form (IV) is more suitable due to its asymptotic flatness at $%
r=\infty $. The zero mass limit is mathematically forbidden in class II
while the class IV solution leads to a trivially flat spacetime in this
limit. In the string theory, massive wormholes exist in the three
corresponding classes of solutions(I,II,IV). A remarkable result is that
zero mass wormholes do {\it not} exist in the string theory at all, at least
within the solution sets considered. A physical reason for this could be
that the gravitational self-energy becomes very large. In this sense, the
non-existence of zero mass wormholes is a result of the violation of SEP in
the string frame. Stable zero mass wormholes that exemplify Wheeler's
\textquotedblleft charge without charge" discussed by Armend\'{a}riz-Pic\'{o}%
n [14] can exist only in EMS class I, and not in other EMS classes, as shown
in this paper. From the analysis in Sec.3, one can now discern the
underlying physical reason: The scalar mass $M_{S}$ does not appear in the
metric (8) so that the gravitational self energy is negligible. Only the EMS
class I solution has this very desirable property.

(3) The motion of a test particle endowed with an infinitesimal
mass and scalar charge has been investigated in both EMS and
string theory following an approach by Buchdahl [22] who used the
Pleba\'{n}ski-Sawicki theorem. The approach is based on the idea
that the test charge responds directly to the scalar field and not
only indirectly via the metric. The metric expansions in the case
of only indirect response have been demonstrated in Eqs.(14) and
(29). If both direct and indirect responses are taken into
account, then the metrics expand like Eqs. (56) and (63). An
interesting result in the case of extremal $M=0$ EMS environment
[action (5)] is that the scalar-scalar interactions produce a
hypothetical scalar \textquotedblleft mass" that can confine test
charges. Similar, but not quite the same, effects occur also in
the string theory described by action (1).

As a mere curiosity, one can speculate a possible cosmological
implication of this phenomenon. Will and Steinhardt [35]
conjectured that an inflation induced oscillation of a massive
gravitational scalar field could account for the \textquotedblleft
missing mass" required to close the universe. The scalar-scalar
interaction at a classical level as considered here could provide
a possible mechanism for the production of the missing mass in the
universe, if one is prepared to allow a violation of WEP. Ordinary
neutral particles does not respond to the scalar mass $M^{\prime
}$ (since the Keplerian mass of the configuration $M=m\beta =0$)
but $M^{\prime }$does curve the local spacetime by way of the
metric (60)-(61). One could think of zero Kepler mass (but
$M^{\prime }\neq 0$) microscopic wormholes populating the universe
and the contributions from $M^{\prime }$ leading to the closure of
the visible universe. However, it is stressed that resolving the
missing mass issue is not the main purpose of the present paper as
the problem involves several other different considerations.

\section{Appendix}

The equivalence of notations of A-P with those in the present paper can be
readily achieved by the identifications: $M_{our} \to m_{A - P} ,\;m_{our}
\to \eta _{A - P} ,\;\sigma _{our} \to q_{A - P}$. The expression $m^2 =\eta
^2 - q^2$ used in Ref. [14] with the assumption $q^2 <0$ is identical to our
Eq. (20) above. It should also be pointed out that the solutions (8) and (9)
reduce, under the radial transformation $r = \rho \left( {1 + \frac{m}{{%
2\rho }}} \right)^2$ and in the A-P notation, to the Janis-Newman-Winnicour
(JNW) form:
$$
ds^2 =\left( {1-\frac{{2\eta }}{\rho }}\right)^{m/\eta } dt^2 -\left( {1-%
\frac{{2\eta }}{\rho }} \right)^{ - m/\eta } d\rho^2 - \left( {1-\frac{{2\eta }%
}{\rho }}\right)^{1 - m/\eta } \rho ^2 d\Omega _2^2 , \eqno(A1)
$$
$$
\varphi (r) = \sqrt {\frac{{1 - \frac{{m^2 }}{{\eta ^2 }}}}{{2\alpha }}} \ln
\left( {1 - \frac{{2\eta }}{r}} \right). \eqno(A2)
$$
For more details, see Ref. [34]. The metric (A1) above is
precisely the Eq.
(21) in A-P [14]. Now, impose the condition of zero total mass, viz., $%
M_{our} \to m_{A - P} = 0$ on (A1). Redefining $\rho =l$, one gets, from
(A1)
$$
ds^2 = dt^2 - dl^2 - \left( {l^2 - 2\eta l} \right)d\Omega _2^2 , \eqno(A3)
$$
$$
\varphi =\frac{1}{2}\ln \left( {1 -\frac{{2\eta }}{l}}\right). \eqno(A4)
$$
Whatever be the nature or sign of $\eta$ the minimum surface area is zero
that occurs either at $l=0$ or at $l=2\eta $ and the scalar field either
blows up or becomes undefined at those values. Thus this form of metric is
not suitable since it represents a naked singularity at $l=0$ or $l=2\eta $.
It should be noted that the solution (8) and (9), which represents the {\it %
same} solution as the one in (A1) and (A2) but only rewritten in isotropic
form, also exhibits a globally strong naked singularity at $r=m/2$. (Visser
[12] called such wormholes ``diseased".) What is interesting is that, only
in the zero mass limit, the disease disappears in one but persists in the
other coordinate system.

\section*{Acknowledgments}

One of us (KKN) wishes to thank Professor Zhong-Can Ouyang for
providing hospitality and excellent working facilities at ITP,
CAS. Unstinted assistance from Mr. Sun Liqun is gratefully
acknowledged. This work is supported in part by the TWAS-UNESCO
program of ICTP, Italy, and as well as by National Basic Research
Program of China under Grant No. 2003CB716300 and by NNSFC under
Grant No.10175070.

\end{document}